\documentclass[10pt,aps,pra,reprint,nobalancelast,nofootinbib,showpacs,
  superscriptaddress,amsmath,amssymb]{revtex4-1}

\usepackage{amsmath}
\usepackage{amssymb}
\usepackage[retainorgcmds]{IEEEtrantools}
\usepackage{dsfont}
\usepackage[colorlinks,linkcolor=blue,citecolor=blue,urlcolor=blue]{hyperref}
\usepackage{graphicx}
\usepackage{psfrag}

\newcommand{\ket}[1]{\lvert#1\rangle}
\newcommand{\bra}[1]{\langle#1\rvert}
\newcommand{\braket}[2]{\langle#1\vert#2\rangle}
\newcommand{\proj}[1]{\ket{#1}\bra{#1}}
\newcommand{\abs}[1]{\lvert#1\rvert}
\newcommand{\comm}[2]{[#1, #2]}
\newcommand{\hilb}{\mathcal{H}}
\newcommand{\transl}{\mathcal{T}}
\newcommand{\subsup}[2]{_{#1}^{\hphantom{#1}#2}}
\newcommand{\id}{\mathds{1}}
\newcommand{\dd}{\mathrm{d}}

\newcommand{\rE}{\mathrm{E}}
\newcommand{\rF}{\mathrm{F}}
\newcommand{\rO}{\mathrm{O}}

\newcommand{\OPERA}{OPERA--Photonique, CP~194/5, Universit\'e libre de
  Bruxelles, av. F.D. Roosevelt 50, Brussels B-1050, Belgium}
\newcommand{\LIQ}{Laboratoire d'Information Quantique, CP~225, Universit\'e
  libre de Bruxelles, av. F.D. Roosevelt 50, Brussels B-1050, Belgium}
\newcommand{\FEMTO}{D\'epartement d'Optique P. M. Duffieux,
  Institut FEMTO--ST, Unit\'e Mixte de Recherche du CNRS 6174, \\
  Universit\'e de Franche-Comt\'e, route de Gray 16, Besan\c{c}on F-25030,
  France}

\begin{document}

\title{Creating and manipulating entangled optical qubits in the frequency
  domain}
\date{\today}
\author{Laurent Olislager}
\email{lolislag@ulb.ac.be}
\affiliation{\OPERA{}}
\author{Erik Woodhead}
\affiliation{\LIQ{}}
\author{Kien Phan~Huy}
\author{Jean-Marc Merolla}
\affiliation{\FEMTO{}}
\author{Philippe Emplit}
\affiliation{\OPERA{}}
\author{Serge Massar}
\affiliation{\LIQ{}}

\begin{abstract}
  Radio-frequency phase modulation of frequency entangled photons leads to a
  two-photon interference pattern in the frequency domain. In recent
  experiments, the pattern was measured with narrow-band frequency filters
  which select photons belonging to a given frequency bin. Here we show how
  photons can be grouped into even and odd frequencies by using periodic
  frequency filters called interleavers. In our theoretical analysis we show
  how this reduces the high-dimensional photon state to an effective
  two-dimensional state. This is of interest for applications such as quantum
  cryptography or low-dimensional tests of quantum nonlocality. We then
  report an experimental realization of this proposal. The observed
  two-photon interference pattern and violation of the CHSH inequality -- the
  simplest binary-outcome Bell inequality -- are in good agreement with the
  theoretical predictions.
\end{abstract}


\maketitle

\section{Introduction}

Entangled photons are a key resource for quantum information processing and
communication. During the past decades, all degrees of freedom of photons
have been used for entanglement experiments, including polarization
\cite{ref:agr1981,ref:agr1982,ref:km1995}, position and momentum
\cite{ref:rt1990}, angular momentum \cite{ref:mv2001}, and time-energy
\cite{ref:bmm1992,ref:ksc1993,ref:trt1993,ref:tb1998,ref:tb2000,ref:mr2004}.
The latter degree of freedom is particularly interesting for long-distance
quantum communication, as it propagates essentially undisturbed through
optical fibers over large distances. Most experiments exploited the concept
of \emph{time bins} originally proposed in \cite{ref:f1989,ref:er1992}, in
which the photons are detected at discrete times. Recently we have introduced
the concept of \emph{frequency bins}, in which the photons are detected
within discrete frequency intervals \cite{ref:oc2010,ref:om2012}. The latter
works are based on earlier works in which the frequency degree of freedom was
used to code information in attenuated coherent pulses for quantum key
distribution applications \cite{ref:mm1999a,ref:mm1999b,ref:bm2007}.

The advantages of frequency-bin entanglement are that it can be manipulated
and measured using standard telecommunication components such as
electro-optic phase modulators and narrow-band fiber Bragg gratings, that raw
visibilities in excess of 99\% can readily be obtained (comparable to the
highest visibilities obtained using other photonic degrees of freedom), that
high-dimensional quantum states can be manipulated (dimension as high as 11
easily obtained), and that no interferometric stabilization is required over
laboratory distance scales (meters of optical fibers).

In the experiments \cite{ref:oc2010,ref:om2012}, electro-optic phase
modulators generated a high-dimensional frequency interference pattern which
was observed with narrow-band frequency filters, each selecting a given
frequency bin. While the high dimensionality of the entangled state can be
beneficial in some quantum information applications, it is sometimes
desirable to work with well-known two-dimensional states for which most
quantum communication protocols, such as the BB84 key distribution scheme
\cite{ref:bb1984}, are designed. In addition, when the states are
two-dimensional, it is easier to access all measurement outcomes
simultaneously since only four detectors are needed, which is better suited
for test of the CHSH Bell inequality \cite{ref:ch1969}. (By contrast, in our
earlier work \cite{ref:oc2010,ref:om2012}, the reported violation of the CH74
inequality \cite{ref:ch1974} on a higher-dimensional frequency entangled
state was based on a simplifying assumption on the marginal statistics that
was not tested directly.)

Here we show how to define, manipulate and measure effective two-dimensional
states in the frequency domain. The key idea is to use as measurement device
a periodic filter that selects two sets of frequency bins, those with
\emph{even} and \emph{odd} frequencies respectively. This implements a
coarse-grained measurement that projects onto two orthogonal subspaces. Such
periodic frequency filters are standard components in the telecommunication
industry, known as \emph{interleavers}. With this approach, we observe a
two-dimensional two-photon interference in the frequency domain and violation
of the CHSH Bell inequality \cite{ref:ch1969}. This is realized by
simultaneously measuring all coincidence probabilities (no further assumption
is needed for the Bell test, contrary to \cite{ref:oc2010,ref:om2012}).

A further interest of the present approach is that it allows one in principle
to manipulate and measure frequency entangled photons that are produced by a
broadband source with low spectral brightness, as the interleavers that
separate the even and odd frequencies act over a very broad bandwidth. This
however requires dispersion compensation, as otherwise photons with
different detunings exhibit different interference patterns that average to
zero over the bandwidth of the photons. 

The paper is divided into two main parts. In the first part we describe
theoretically how effective qubits corresponding to even and odd frequencies
can be introduced, and how electro-optic phase modulators realize rotations
in this two-dimensional space. We compute how the two-photon correlations
depend on the amplitude and phase of the radio-frequency signals driving the
phase modulators. From these expressions we show that the maximum possible
violation of the CHSH inequality is 2.566. In the second part we describe our
experimental setup. We report two-photon interference patterns in good
agreement with the theoretical predictions, and report a violation of the
CHSH inequality of $2.334 \pm 0.008$. The reader principally interested in
the experiment may skip directly to section~\ref{sec:exp}.

\section{Outline of the experiment\label{sec:outline}}

\begin{figure}[t]
  \centering
  \includegraphics[width=.99\columnwidth]{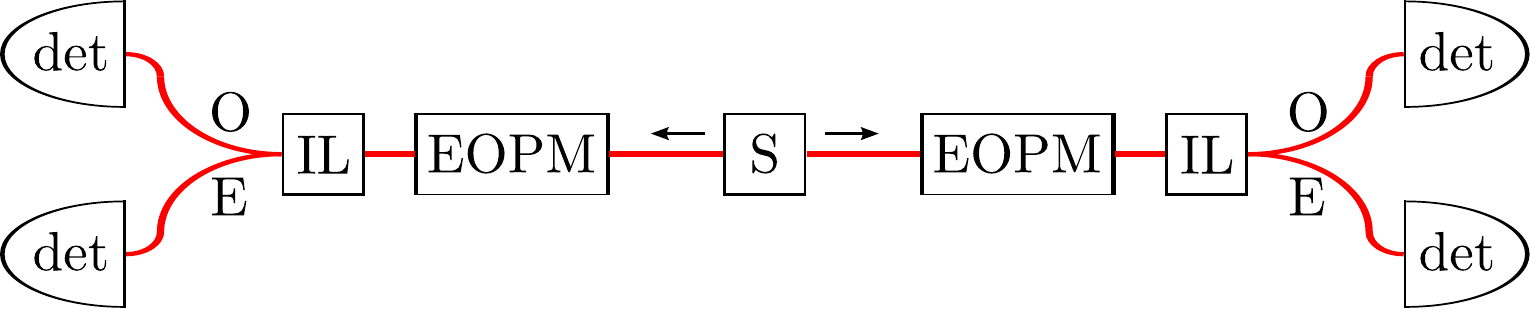}
  \caption{Simple schematic of the experiment. The source (S) produces
    frequency entangled photons. Two electro-optic phase modulators (EOPM)
    driven by radio-frequency signals with identical frequency $\Omega$ but
    different amplitudes and phases, $(a, \alpha)$ and $(b, \beta)$, realize
    interference in the frequency domain. Interleavers (IL) send the even (E)
    and odd (O) frequency bins to separate single-photon detectors
    (det). \label{fig:outline}}
\end{figure}

Our experiment is schematized in Fig.~\ref{fig:outline}. It is based on three
components that we briefly describe:

1) A source S produces the frequency entangled state
\begin{IEEEeqnarray}{rCl}
  \ket{\Psi} &=& \int \dd \omega f(\omega) \ket{\omega_0 + \omega}
  \ket{\omega_0 - \omega} \IEEEnonumber \\
  &\simeq& \int \dd \omega \ket{\omega_0 + \omega} \ket{\omega_0 - \omega} \,,
\end{IEEEeqnarray}
where $f(\omega)$ characterizes the two-photon bandwidth. Because $f$ varies
slowly with frequency, for the theoretical analysis it is often useful to
approximate it as constant as in the second line.

2) A phase modulator driven by a radio-frequency signal $v \cos(\Omega t -
\gamma)$ with adjustable amplitude $v$ and phase $\gamma$ realizes the
unitary transformation
\begin{equation}
  \label{eq:pm_act_def}
  \ket{\omega} \mapsto \sum_{p \in \mathbb{Z}} J_p(c) e^{ip(\gamma-\pi/2)}
  \ket{\omega + p \Omega} \,,
\end{equation}
$J_{p}(c)$ being the $p$th-order Bessel function of the first kind, with
normalized amplitude $c = \pi v / V_\pi$ where $V_{\pi}$ characterizes the
response of the modulator.

3) An interleaver is a component used in the telecommunication industry that
separates the frequencies centered on $\omega_{0} + 2 n \Omega$ from those
centered on $\omega_{0} + (2n + 1) \Omega$, where $\omega_{0}$ is a fixed
offset, and $n \in \mathbb{Z}$. We shall use interleavers as components that
allow the measurement of even and odd frequencies. If we follow the interleaver
by single-photon detectors, then a click of one of the detectors corresponds
to the projection onto one of the two operators:
\begin{IEEEeqnarray}{rCl}
  \label{eq:proj_eo_def}
  \Pi_{\rE} &=& \int_{-\Omega}^{+\Omega} \dd \omega g(\omega)
  \sum_{n} \Pi_{\omega_{0} + \omega + 2n \Omega} \,, \\
  \Pi_{\rO} &=& \int_{-\Omega}^{+\Omega} \dd \omega g(\omega)
  \sum_{n} \Pi_{\omega_{0} + \omega + (2n + 1) \Omega} \,,
\end{IEEEeqnarray}
where $\Pi_{\omega} = \proj{\omega}$ is the projector onto the frequency
state $\ket{\omega}$ and $g(\omega)$ is a function characteristic of the
interleaver which is maximal in the vicinity of $\omega = 0$ and very small
when $\abs{\omega} > \Omega/2$. Examples of transmission spectra of
interleavers can be found in Fig.~\ref{fig:filters}.

The experiment consists of preparing the state, sending Alice's and Bob's
photons through phase modulators driven by radio-frequency signals with
identical frequency $\Omega$ but different amplitudes and phases, ($a,
\alpha$) and ($b, \beta$), and finally determining whether the frequency is
even or odd by passing the photon through interleavers and then sending the
output to single-photon detectors.

\section{Theoretical analysis\label{sec:theory}}

\subsection{Discrete and offset space}

Because the phase modulator shifts the frequency by $\Omega$ and the
interleaver is sensitive only to frequency modulo $2 \Omega$, it is
convenient to rewrite the state as:
\begin{IEEEeqnarray}{rCll}
  \label{eq:psi_sep}
  \ket{\Psi} &=& \int_{-\Omega/2}^{+\Omega/2} \dd \omega \sum_{n=-\infty}^{+\infty}
  & f(\omega + n\Omega) \IEEEnonumber \\
  &&& \times\: \ket{\omega_{0} + \omega + n \Omega}
  \ket{\omega_{0} - \omega - n \Omega} \IEEEnonumber \\
  &\simeq& \IEEEeqnarraymulticol{2}{l}{
    \int_{-\Omega/2}^{+\Omega/2} \dd \omega \sum_{n=-\infty}^{+\infty}
    f_{n} \ket{n, \omega} \ket{-n, -\omega}} \IEEEnonumber \\ 
  &=& \IEEEeqnarraymulticol{2}{l}{
    \sum_{n \in \mathbb{Z}} f_{n} \ket{n} \ket{-n}
    \otimes \int_{-\Omega/2}^{\Omega/2} \dd \omega \ket{\omega}
    \ket{-\omega}} \IEEEnonumber \\
  &=& \IEEEeqnarraymulticol{2}{l}{
    \ket{\Psi_{\Pi}} \otimes \ket{\Psi_{\mathrm{off}}} \,,}
\end{IEEEeqnarray}
where we suppose that $f$ varies slowly so that we can neglect the dependence
on $\omega$: $f(\omega + n \Omega) \simeq f_{n}$. The identification $\ket{n}
\otimes \ket{\omega} = \ket{n, \omega} = \ket{\omega_{0} + n\Omega + \omega}$
defines a factorization $\hilb_{\rF} = \hilb_{\Pi} \otimes
\hilb_{\mathrm{off}}$ of the Hilbert space $\hilb_{\rF}$ of frequency states
into separate ``discrete'' and ``offset'' spaces, $\hilb_{\Pi}$ and
$\hilb_{\mathrm{off}}$, respectively, with respect to which the source state
$\ket{\Psi}$ is (approximately) separable. We adopt the normalization
\begin{IEEEeqnarray}{rCl}
  \braket{m}{n} &=& \delta_{mn} \,, \\
  \braket{\omega'}{\omega} &=& \delta(\omega' - \omega) \,.
\end{IEEEeqnarray}

We factorize the projections \eqref{eq:proj_eo_def} in a similar manner,
\begin{IEEEeqnarray}{rCl}
  \Pi_{\rE} &\simeq& \sum_{n} \proj{2n}
  \otimes \int_{-\Omega/2}^{\Omega/2} \dd \omega
  g(\omega) \proj{\omega} \,, \IEEEnonumber \\
  \Pi_{\rO} &\simeq& \sum_{n} \proj{2n + 1}
  \otimes \int_{-\Omega/2}^{\Omega/2} \dd \omega g(\omega) \proj{\omega} \,,
\end{IEEEeqnarray}
where our requirement that $g(\omega)$ is negligible for $\abs{\omega} >
\Omega/2$ justifies restricting the integration over $\omega$ to the range
$[-\Omega/2, \Omega/2]$. Finally, in this representation, the action
\eqref{eq:pm_act_def} of a phase modulator takes the expression
\begin{IEEEeqnarray}{rCl}
  \ket{n} \otimes \ket{\omega}
  &\mapsto& \sum_{p \in \mathbb{Z}} J_{p}(c) e^{ip(\gamma - \pi/2)}
  \ket{n + p} \otimes \ket{\omega} \IEEEnonumber \\
  &&= \bigl( U(c, \gamma) \otimes \id_{\mathrm{off}} \bigr)
  \ket{n} \otimes \ket{\omega} \,,
\end{IEEEeqnarray}
with the unitary transformation $U(c, \gamma)$ acting only on $\hilb_{\Pi}$,
and $\id_{\mathrm{off}}$ is the identity in the ``offset'' space.

We see explicitly, then, that the description of the relevant part of our
setup is entirely contained in the discrete space $\hilb_{\Pi}$: indeed, the
offset frequency $\omega$ only affects the probability of response of the
interleaver via the factor $\abs{g(\omega)}^2$, and is otherwise never
measured or recorded in the course of the experiment. Consequently, we will
restrict the remainder of the theoretical analysis to this space. Note that
the factorization and isolation of the discrete space $\hilb_{\Pi}$ detailed
here formalizes the concept of ``frequency bin'' previously used in
\cite{ref:oc2010,ref:om2012}.

\subsection{Phase states and effective qubits}

The effective qubits manipulated in our setup are made explicit when we
express the source state $\ket{\Psi_{\Pi}}$ and actions of the phase
modulators and interleavers in the basis of \emph{even} and \emph{odd phase
  states}. These states can be derived from our setup's symmetries with
respect to translations of frequency bins. Formally, let us denote
\begin{equation}
  \transl_{k} \colon \ket{n} \mapsto \ket{n + k}
\end{equation}
the (unitary) operation consisting of translation in the frequency domain by
$k$ frequency bins. The phase modulator and interleaver actions are symmetric
with respect to translations by $k$ and $2k$, respectively, in the sense that
\begin{equation}
  \comm{U(c, \gamma)}{\transl_{k}} = 0 \,, \quad k \in \mathbb{Z} \,,
\end{equation}
and
\begin{equation}
  \comm{\Pi_{\rE}}{\transl_{k}}
  = \comm{\Pi_{\rO}}{\transl_{k}} = 0 \,, \quad k \in 2 \mathbb{Z} \,,
\end{equation}
while, using that the amplitude $f_{n}$ varies slowly, the source state has
the approximate symmetry
\begin{equation}
  \transl_{k} \otimes \transl_{-k} \ket{\Psi_{\Pi}} \simeq \ket{\Psi_{\Pi}} \,.
\end{equation}
Consequently, the phase modulators and source will share eigenstates with the
$\transl_{1}$ operator, while the interleaver action eigenstates will
coincide with those of $\transl_{2}$.

A full set of eigenstates of the $\transl_{1}$ operator is given by the phase
states, which we define by
\begin{equation}
  \ket{\varphi} = \frac{1}{\sqrt{2 \pi}}
  \sum_{n \in \mathbb{Z}} e^{i n \varphi} \ket{n} \,,
\end{equation}
such that $\transl_{k} \ket{\varphi} = e^{-i k \varphi} \ket{\varphi}$. The
inverse of this expression is given by
\begin{equation}
  \ket{n} = \frac{1}{\sqrt{2 \pi}}
  \int_{-\pi}^{\pi} \dd \varphi e^{-i n \varphi} \ket{\varphi} \,.
\end{equation}
For the $\transl_{2}$ operator, a complete basis of eigenstates that will
prove convenient is given by the even and odd phase states
\begin{IEEEeqnarray}{rCl}
  \ket{\varphi}_{\rE} &=& \frac{1}{\sqrt{\pi}}
  \sum_{n \in 2 \mathbb{Z}} e^{i n \varphi} \ket{n} \,, \\
  \ket{\varphi}_{\rO} &=& \frac{1}{\sqrt{\pi}}
  \sum_{n \in 2 \mathbb{Z} + 1} e^{i n \varphi} \ket{n} \,,
\end{IEEEeqnarray}
with $\transl_{2} \ket{\varphi}_{\rE} = e^{-2 i \varphi} \ket{\varphi}_{\rE}$
and $\transl_{2} \ket{\varphi}_{\rO} = e^{-2 i \varphi}
\ket{\varphi}_{\rO}$. Note that
\begin{IEEEeqnarray}{rCl}
  \label{eq:ph_eo_rel}
  \ket{\varphi} &=& \frac{1}{\sqrt{2}} \bigl[ \ket{\varphi}_{\rE}
  + \ket{\varphi}_{\rO} \bigr] \,, \IEEEnonumber \\
  \ket{\varphi + \pi} &=& \frac{1}{\sqrt{2}} \bigl[ \ket{\varphi}_{\rE}
  - \ket{\varphi}_{\rO} \bigr] \,.
\end{IEEEeqnarray}
In terms of these states, the even and odd projection operators (restricted
to the discrete space) take the expressions
\begin{IEEEeqnarray}{rCl}
  \label{eq:projE}
  \Pi_{\rE} &=& \int_{0}^{\pi} \dd \varphi
  \ket{\varphi}_{\rE} \bra{\varphi}_{\rE} \,, \\
  \label{eq:projO}
  \Pi_{\rO} &=& \int_{0}^{\pi} \dd \varphi
  \ket{\varphi}_{\rO} \bra{\varphi}_{\rO} \,,
\end{IEEEeqnarray}
and the entangled source state can be rewritten
\begin{IEEEeqnarray}{rCl}
  \label{eq:psipi}
  \ket{\Psi_{\Pi}} &\simeq& \frac{1}{\sqrt{N}}
  \sum_{n} \ket{n} \ket{-n} \IEEEnonumber \\
  &=& \frac{1}{\sqrt{N}} \int_{-\pi}^{\pi} \dd \varphi
  \ket{\varphi} \ket{\varphi} \IEEEnonumber \\
  &=& \frac{1}{\sqrt{N}} \int_{0}^{\pi} \dd \varphi \bigl(
  \ket{\varphi}_{\rE} \ket{\varphi}_{\rE}
  + \ket{\varphi}_{\rO} \ket{\varphi}_{\rO} \bigr) \,,
\end{IEEEeqnarray}
where we idealize $\ket{\Psi_{\Pi}}$ as an infinite sum, and $N$ is a
normalization constant symbolically representing the number of frequency bins
over which $f_n$ is non zero, and formally equal to $2 \pi \delta(0)$ (see
also the discussion of normalization in \cite{ref:oc2010}). To obtain the
second line, we used that
\begin{equation}
  \sum_{n \in \mathbb{Z}} e^{i n \theta}
  = 2 \pi \sum_{k \in \mathbb{Z}} \delta(\theta - 2 \pi k) \,.
\end{equation}
The action of a phase modulator on a phase state is found to be
\begin{IEEEeqnarray}{rCl}
  \label{eq:PhmodPhasestate}
  U(c, \gamma) \ket{\varphi} &=& \frac{1}{\sqrt{2 \pi}}
  \sum_{m} e^{i m \varphi} \sum_{p} J_{p}(c)
  e^{i p (\gamma - \pi/2)} \ket{m + p} \IEEEnonumber \\
  &=& \sum_{p} J_{p}(c) e^{i p (\gamma - \varphi - \pi/2)}
  \frac{1}{\sqrt{2 \pi}} \sum_{n} e^{i n \varphi} \ket{n} \IEEEnonumber \\
  &=& e^{-i c \cos(\gamma - \varphi)} \ket{\varphi} \,,
\end{IEEEeqnarray}
where, to obtain the last line, we used a version of the Jacobi-Anger
expansion \cite{ref:cbp2008}:
\begin{equation}
  e^{- i c \cos(\theta)} = \sum_{n} J_{n}(c) e^{i n (\theta - \pi/2)} \,.
\end{equation}

Using Eqs.~\eqref{eq:ph_eo_rel} and \eqref{eq:PhmodPhasestate} we readily find
\begin{IEEEeqnarray}{rCl}
  U(c, \gamma) \ket{\varphi}_{\rE} &=& \cos(\theta) \ket{\varphi}_{\rE}
  - i \sin(\theta) \ket{\varphi}_{\rO} \,, \IEEEnonumber \\
  U(c, \gamma) \ket{\varphi}_{\rO} &=& - i \sin(\theta) \ket{\varphi}_{\rE}
  + \cos(\theta) \ket{\varphi}_{\rO} \,,
\end{IEEEeqnarray}
where we have set $\theta = c \cos(\gamma - \varphi)$. For a fixed phase
$\varphi$, we see that, varying the modulation parameters $c$ and $\gamma$,
we can implement a $\sigma_{x}$ rotation of any desired angle between the
even and odd phase states.

The above construction thus shows how to define effective qubits
$\{\ket{\varphi}_{\rE}, \ket{\varphi}_{\rO}\}$ in the frequency domain, and
how phase modulators realize $\sigma_x$ rotations on the effective
qubits. However the angle of the rotation depends on the phase $\varphi$ of
the effective qubit. This phase is not experimentally accessible. As the
entangled state Eq.~\eqref{eq:psipi} is given by an integral over $\varphi$,
this will imply a modified interference pattern with reduced visibility. In
the next section we quantify this and show that the proposed experiment
allows violation of the CHSH inequality.

\subsection{Two-photon interference pattern}

Modulating each arm of our setup with the modulation parameters $A = (a,
\alpha)$ and $B = (b, \beta)$ transforms the initial source state to
\begin{IEEEeqnarray}{rCll}
  \ket{\Psi_{AB}} &=& \frac{1}{\sqrt{N}}
  \int_{0}^{\pi} \dd \varphi \bigl\{ &
  \cos(\theta_{A} + \theta_{B}) \ket{\phi^{+}_{\varphi}} \IEEEnonumber \\
  &&&-\, i \sin(\theta_{A} + \theta_{B}) \ket{\psi^{+}_{\varphi}} \bigr\} \,,
\end{IEEEeqnarray}
where we have set $\theta_{A} \equiv \theta_{A}(\varphi) = a \cos(\varphi -
\alpha)$ and $\theta_{B} \equiv \theta_{B}(\varphi) = b \cos(\varphi -
\beta)$, and
\begin{IEEEeqnarray}{rCl}
  \ket{\phi^{+}_{\varphi}} &=& \ket{\varphi}_{\rE} \ket{\varphi}_{\rE}
  + \ket{\varphi}_{\rO} \ket{\varphi}_{\rO} \,, \\
  \ket{\psi^{+}_{\varphi}} &=& \ket{\varphi}_{\rE} \ket{\varphi}_{\rO}
  + \ket{\varphi}_{\rO} \ket{\varphi}_{\rE} \,.
\end{IEEEeqnarray}
Via elementary trigonometric identities, we have
\begin{equation}
  \theta_{A}(\varphi) + \theta_{B}(\varphi)
  = D \cos(\varphi - \Delta) \equiv \theta_{AB}(\varphi) \,,
\end{equation}
with
\begin{equation}
  D^{2} = a^{2} + b^{2} + 2 ab \cos(\alpha - \beta)
\end{equation}
and
\begin{equation}
  \tan(\Delta) = \frac{a \sin(\alpha) + b \sin(\beta)}
  {a \cos(\alpha) + b \cos(\beta)} \,.
\end{equation}

The probability of jointly detecting two photons in even frequency bins is
then given by
\begin{IEEEeqnarray}{rCll}
  \label{eq:P_ee}
  P(\rE, \rE) &=& \IEEEeqnarraymulticol{2}{l}{
    \bra{\Psi_{AB}} \Pi_{\rE} \otimes \Pi_{\rE} \ket{\Psi_{AB}}}
  \IEEEnonumber \\
  &=& \frac{1}{N} \int_{0}^{\pi} \dd \varphi' \int_{0}^{\pi} \dd \varphi \,
  &\cos \bigl( \theta_{AB}(\varphi') \bigr)
  \cos \bigl( \theta_{AB}(\varphi) \bigr) \IEEEnonumber \\
  &&& \times\: \bra{\varphi'}_{\rE} \Pi_{\rE} \ket{\varphi}_{\rE}
  \bra{\varphi'}_{\rE} \Pi_{\rE} \ket{\varphi}_{\rE} \IEEEnonumber \\
  &=& \IEEEeqnarraymulticol{2}{l}{
    \frac{\delta(0)}{N} \int_{0}^{\pi} \dd \varphi
    \cos \bigl( \theta_{AB}(\varphi) \bigr)^{2}} \IEEEnonumber \\
  &=& \IEEEeqnarraymulticol{2}{l}{
    \frac{1}{4} + \frac{1}{4 \pi} \int_{0}^{\pi} \dd \varphi
    \cos \bigl( 2 D \cos(\varphi - \Delta) \bigr)} \IEEEnonumber \\
  &=& \IEEEeqnarraymulticol{2}{l}{
    \frac{1}{4} \bigl[ 1 + J_{0}(2D) \bigr] \,.}
\end{IEEEeqnarray}
To reach the last line, we used the integral expression
\begin{equation}
  J_{0}(x) = \frac{1}{\pi} \int_{0}^{\pi} \dd t \cos \bigl( x \sin(t) \bigr)
\end{equation}
for the zeroth Bessel function of the fist kind, and that the function $t
\mapsto \cos \bigl(x \sin(t) \bigr)$ is $\pi$-periodic in $t$. We similarly
find
\begin{equation}
  \label{eq:P_oo}
  P(\rO, \rO) = \frac{1}{4} \bigl[ 1 + J_{0}(2D) \bigr]
\end{equation}
and
\begin{equation}
  \label{eq:P_eo_P_oe}
  P(\rE, \rO) = P(\rO, \rE) = \frac{1}{4} \bigl[ 1 - J_{0}(2D) \bigr] \,.
\end{equation}

Note that $P(\rE, \rE)$ and $P(\rO, \rO)$ never vanish, whereas $P(\rE, \rO)$
and $P(\rO, \rE)$ vanish whenever $D = 0$, which occurs whenever $a = b$ and
$\alpha - \beta = \pi$. Because of the average over $\varphi$, the
interference pattern differs from the traditional sine-squared function.

\subsection{Maximal violation of the CHSH inequality\label{sec:maxCHSH}}

The main result of our experiment consists of an estimation of the CHSH
expression
\begin{equation}
  \label{eq:CHSHexp}
  S = E(A_{0} B_{0}) + E(A_{0} B_{1}) + E(A_{1} B_{0}) - E(A_{1} B_{1}) \,,
\end{equation}
where $A_{i} \equiv (a_{i}, \alpha_{i})$ and $B_{j} \equiv (b_{j},
\beta_{j})$ denote choices of modulation amplitudes and phases,
\begin{IEEEeqnarray}{rCl}
  E(A_{i} B_{j}) &=& P_{ij}(\rE, \rE) - P_{ij}(\rE, \rO) \IEEEnonumber \\
  &&-\, P_{ij}(\rO, \rE) + P_{ij}(\rO, \rO) \,,
\end{IEEEeqnarray}
and, e.g., $P_{ij}(\rE, \rE)$ is the probability of detecting two photons of
even parity following modulation with the parameters $A_{i}$ and
$B_{j}$. Using Eqs.~\eqref{eq:P_ee}, \eqref{eq:P_oo}, and
\eqref{eq:P_eo_P_oe}, the CHSH correlator is given by
\begin{equation}
  S = J_{0}(2 D_{00}) + J_{0}(2 D_{01}) + J_{0}(2 D_{10}) - J_{0}(2 D_{11})
\end{equation}
with
\begin{equation}
  D\subsup{ij}{2} = a\subsup{i}{2} + b\subsup{j}{2}
  + 2 a_{i} b_{j} \cos(\alpha_{i} - \beta_{j}) \,.
\end{equation}

Following reasoning similar to that in \cite{ref:om2012}, we find that $S$ is
maximized by choosing modulation amplitudes and phases in such a way that
$D_{00} = D_{01} = D_{10} = D_{11} / 3$. This is achieved with phases given
by $\alpha_{0} = \beta_{0} = \gamma$ and $\alpha_{1} = \beta_{1} = \gamma +
\pi$ for some $\gamma$, and modulation amplitudes satisfying $a_{0} = b_{0} =
c$ and $a_{1} = b_{1} = 3 c$. We take, for $c$, the value that maximizes
\begin{equation}
  S(c) = 3 J_{0}(4 c) - J_{0}(12 c) \,,
\end{equation}
which is readily found numerically. The optimal modulation amplitudes are
found this way to be $a_{0} = b_{0} = 0.2318$ and $a_{1} = b_{1} =
0.6955$. With these parameters, the CHSH correlator attains a maximal
theoretical value of $S = 2.566$, thereby demonstrating that even though the
interference is not perfect, a significant violation of the CHSH inequality
is possible in this experiment.

\begin{figure}[!h]
  \centering
  \includegraphics[width=.87\columnwidth]{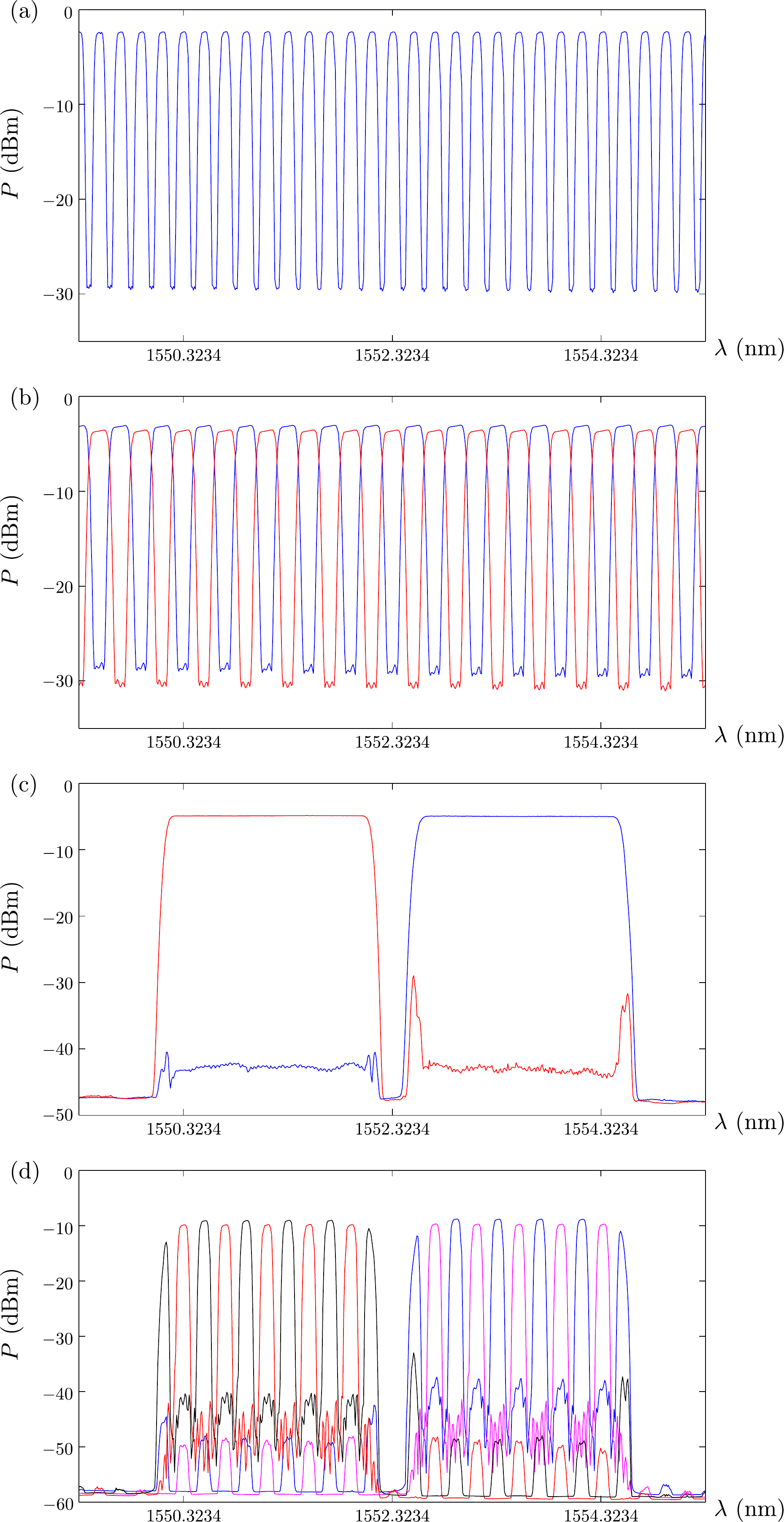}
  \caption{Spectrum of the filters used in the experiment. Since $\lambda_{p}
    = 776.1617\,$nm, 1552.3234\,nm corresponds to the degeneracy frequency
    $\omega_{0}$. From top to bottom: a) Output of the 12.5--25
    interleaver. b) Even (red curve) and odd (blue curve) outputs of a 25--50
    interleaver. c) Programmable WaveShaper filter; photons belonging to the
    red (resp.\ blue) output are sent to Alice (resp.\ Bob). d) Spectrum
    obtained when cascading 12.5--25 interleaver, WaveShaper and 25--50
    interleaver; red curve: Alice, even; black curve: Alice, odd; magenta
    curve: Bob, even; blue curve: Bob, odd. Note that whereas the outputs of
    the 25-50 interleavers (panel b) have $\approx 25\,$dB extinction at the
    center of each pass band, they only have $\approx 3\,$dB extinction at
    the edges of the band (where the red and blue curves cross). Hence
    photons at the edges of the pass bands have quite high and equal
    probabilities to exit the even and odd ports, which would result in
    an important decrease of visibility of interference if the 25--50
    interleavers were used alone. The spectra in panel d show that upon using
    the inital 12.5--25 interleaver (depicted in panel a) that removes the
    photons at the edges of the pass bands, the even and odd outputs are now
    separated by 25\,dB over the whole frequency band. \label{fig:filters}}
\end{figure}

\section{Experimental setup \label{sec:exp}}

\begin{figure}[t]
  \centering
  \includegraphics[width=.99\columnwidth]{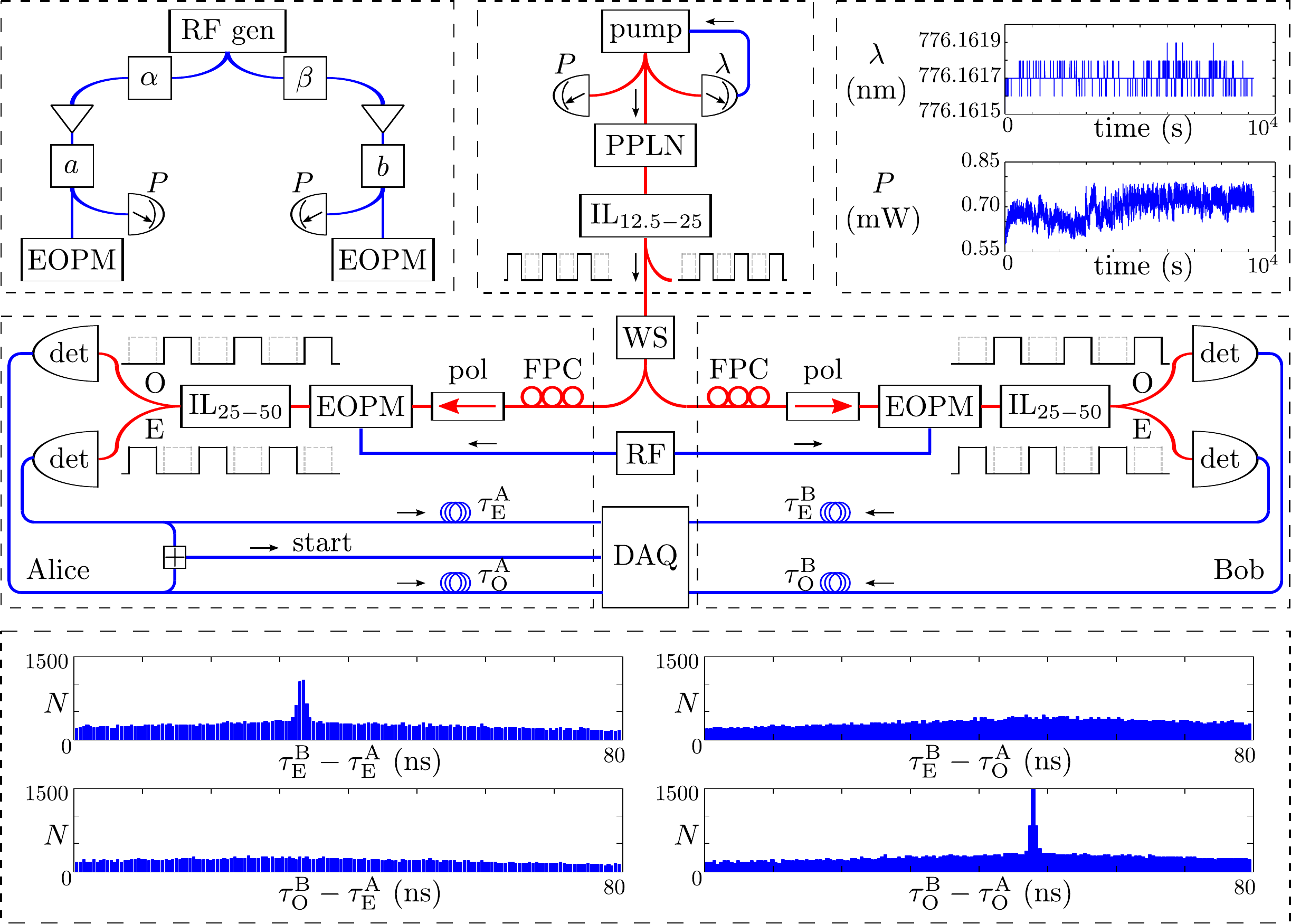}
  \caption{Experimental setup. Red links are optical fibers and blue links
    are electronic connections. A pump laser whose power $P$ and wavelength
    $\lambda$ are continuously monitored is directed through a periodically
    poled lithium niobate waveguide (PPLN). (Upper right: measurements of $P$
    and $\lambda$ during the experiment; although $P$ fluctuates, a
    retroaction loop acting on the piezoelectric element of the external
    cavity diode laser ensures that $\lambda$ is constant). Generated photon
    pairs pass through a 12.5--25\,GHz interleaver (IL$_{12.5-25}$) and a
    programmable filter (WS) which separates signal and idler photons,
    respectively sent to Alice and Bob. On each arm, photons pass through a
    fiber polarization controller (FPC), a polarizer (pol) and an
    electro-optic phase modulator (EOPM) driven by a radio-frequency (RF)
    signal. (Upper left: schematic of the RF circuit. RF signals are
    generated by a 25-GHz RF generator whose power is split between Alice and
    Bob. On each arm, a variable phase shifter, an amplifier and a variable
    attenuator ensure the precise adjustment of phase $\alpha, \beta$ and
    amplitude $a, b$, this last quantity being measured by a powermeter at
    the 10\% output of a directional coupler placed before the
    EOPM. Isolators in the circuit (not shown) ensure that unwanted
    reflections do not distort the values of $a, b, \alpha, \beta$). After a
    25--50\,GHz interleaver (IL$_{25-50}$), single-photon detectors (det)
    record even (E) and odd (O) results. A data acquisition system (DAQ)
    registers detection coincidences and outputs histograms of these
    events. The DAQ is triggered by the arrival of a photon in one of Alice's
    detectors (start signal). The DAQ then records the exact time of arrival
    $\tau^{\mathrm{A},\mathrm{B}}_{\rE,\rO}$ of photons coming from Alice's and Bob's
    detectors. (Bottom: typical results when no phase modulation is applied:
    one observes only EE and OO coincidences; with phase modulation, EO and
    OE coincidences would appear due to two-photon interference in the
    frequency domain.) \label{fig:setup}}
\end{figure}

The details of our experimental setup are depicted in
Fig.~\ref{fig:setup}. It is composed of commercially-available
fiber-pigtailed and opto-electronic components and operates in the
telecommunication C-band.

A continuous laser (Sacher) with power $P \approx 0.7\,$mW and stabilized
wavelength $\lambda_{p} = 776.1617\,$nm pumps a periodically poled lithium
niobate waveguide (HC Photonics), generating the frequency-entangled state
\begin{equation}
  \ket{\Psi} = \int \dd \omega f(\omega) \ket{\omega_{0} + \omega}
  \ket{\omega_{0} - \omega} \,,
\label{eq:psi1}
\end{equation}
with $\omega_{0} / 2 \pi = c / 2\lambda_p=193.125\,$THz and $f(\omega)$
characterizes the two-photon bandwidth (approximately 5\,THz). In what
follows, we relate frequencies to the International Telecommunication Union
DWDM grid in the C-band: multiples of 50\,GHz are said to be on the 50-grid,
multiples of 25\,GHz are on the 25-grid, and other frequencies are off the
grid.

In order to create a nice frequency comb, the photons pass through a 12.5--25
frequency interleaver (Optoplex). The photons whose frequencies belong to
a-few-GHz-wide intervals centered on the 25-grid are collected at the output,
while those centered on intervals with a 12.5\,GHz offset are thrown away
with more than 25\,dB extinction. The reason for using this first filter is
explained in the caption of Fig.~\ref{fig:filters}, where the transmission
spectra of all filters used in the experiment are shown.

The state at the output of this periodic filter can be written as in
Eq.~\eqref{eq:psi_sep}:
\begin{IEEEeqnarray}{rCl}
  \label{eq:psi_sep_pf}
  \ket{\Psi} &=& \sum_{n \in \mathbb{Z}} f_{n} \ket{n} \ket{-n}
  \otimes \int_{-\Omega/2}^{\Omega/2} \dd \omega h(\omega)
  \ket{\omega} \ket{-\omega} \,,
\end{IEEEeqnarray}
where $\Omega = 25$\,GHz and $h(\omega)$ is a function that represents the
effect of the 12.5--25 frequency interleaver (it is maximal around $\omega =
0$ and tends rapidly to zero). The pump is rejected with more than 100\,dB
extinction when taking into account all filters preceding detection.

Photons then pass through a programmable filter (WaveShaper from Finisar)
which is configured to direct photons from bins $n = + (\mathrm{resp.\ }-) 1,
2, 3, 4, 5, 6$ to Alice (resp. Bob). Thus we obtain the state
\begin{equation}
  \label{eq:psi_6bins}
  \ket{\Psi} = \frac{1}{\sqrt{6}} \sum_{n=1}^{6} \ket{n} \ket{-n}
  \otimes \int_{-\Omega/2}^{\Omega/2} \dd \omega h(\omega)
  \ket{\omega} \ket{-\omega} \,,
\end{equation}
where we omit the factors $f_{n}$ on such a reduced bandwidth. The
restriction to only 6 frequency bins is realized so that dispersion can be
neglected. Otherwise, photons in different frequency bins accumulate
different phase shits during propagation through the optical fibers that
deteriorate the two-photon interference pattern. The number of frequency bins
could be increased if dispersion compensation were implemented. Note that
limiting the number of frequency bins will decrease the visibility of the
interference pattern.

On each arm, a polarization controller followed by a polarizer ensures that
the polarization of the photons is aligned with the axis of an electro-optic
phase modulator (EOspace) driven by an adjustable 25-GHz radio-frequency
signal. The radio-frequency architecture shown in the inset of
Fig.~\ref{fig:setup} allows the phase modulation of each photon by
radio-frequency signals $a \cos(\Omega t - \alpha)$, $b \cos(\Omega t -
\beta)$ with independently adjustable amplitude $a, b$ and phase $\alpha,
\beta$.

Finally, the photons are directed to a 25--50 frequency interleaver. One
output collects photons belonging to the 50-grid, i.e.\ frequency bins with
$n$ odd (result O), while the other collects photons remaining from the
25-grid, i.e.\ frequency bins with $n$ even (result E).

Four single-photon detectors (avalanche photodiodes id200 and id201 from
idQuantique, efficiency 10\%, repetition rate 100\,kHz, gate width 100\,ns,
dark-count rates 0.2--0.6 kHz) allow the simultaneous acquisition of EE, EO,
OE and OO coincidences by a data acquisition system (Agilent
Acqiris). Triggered by a detection on Alice's side, it registers the relative
times between detections and outputs histograms of these events.

\section{Experimental results}

Histograms at the bottom of Fig.~\ref{fig:setup} correspond to coincidences
in 0.5\,ns steps measured during half an hour when no phase modulation was
applied. One can see that only EE and OO coincidences are present, as
expected by Eq.~\eqref{eq:psi_6bins}. We note that it is possible to change
the correlations by changing the wavelength of the pump: e.g., when
$\lambda_{p} = 776.1115\,$nm, we measure inverted correlations.

Coincidences are measured at a rate $\approx$ 1.5\,Hz and with a
coincidence-to-accidental ratio $\approx$ 2. These low values are due to the
high losses from pair creation to detection ($\approx 18\,$dB for each
channel), and to the gated operation and high dark-count rates of the
detectors used.

\begin{figure}[t]
  \centering
  \includegraphics[width=.99\columnwidth]{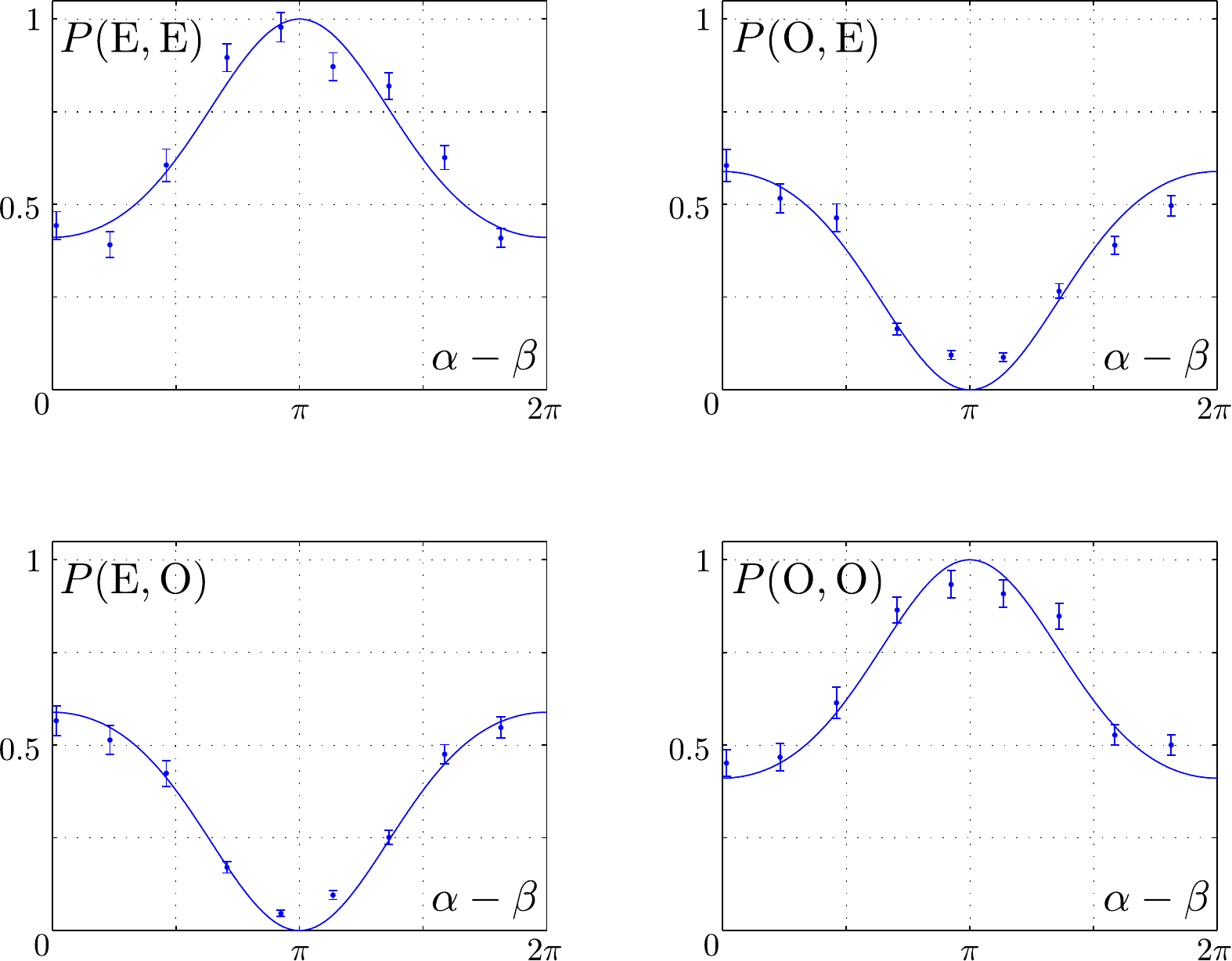}
  \caption{Two-dimensional two-photon interference patterns. Parameters are:
    $a=b=0.6955$, and $\alpha$ is changed with $\beta$ kept constant. Curves
    are theoretical predictions for coincidence probabilities $P(\rE, \rE)$,
    $P(\rE, \rO)$, $P(\rO, \rE)$ and $P(\rO, \rO)$, see Eqs.~\eqref{eq:P_ee},
    \eqref{eq:P_oo}, and \eqref{eq:P_eo_P_oe}. Symbols are experimental
    results: they correspond to the number of coincidences $N(\rE, \rE)$,
    $N(\rE, \rO)$, $N(\rO, \rE)$ and $N(\rO, \rO)$ simultaneously registered
    for each combination of outputs. Note that a normalization based on the
    coincidence rates registered when modulation is off is realized; error
    bars are statistical; background noise of the histograms has been
    subtracted. The net interference visibility (calculated on curves that
    should cancel) is evaluated to be $(85 \pm 5) \%$, depending on the
    combination considered.\label{fig:results}}
\end{figure}

The experimental measurements, some of which are shown and commented in
Fig.~\ref{fig:results}, are in good agreement with the theoretical
predictions, Eqs.~\eqref{eq:P_ee}, \eqref{eq:P_oo}, and
\eqref{eq:P_eo_P_oe}. When $a = b$, the probabilities $P(\rE, \rO)$ and
$P(\rO, \rE)$ should vanish when the phase difference $\alpha - \beta$ is
scanned, which enables one to define the visibility of the interference
fringes as $V = (N_{\text{max}} - N_{\text{min}}) / (N_{\text{max}} +
N_{\text{min}})$, where $N_{\text{max},\text{min}}$ are the net (dark counts
subtracted) maximum and minimum number of counts per unit time. For the value
$a = b = 0.6955$ used in the figure, we measure $V = 90\%$ and $V = 80\%$
depending on which combination, EO or OE, is considered. This limited
visibility is attributed to non-ideal state preparation: limited bandwidth
and dispersion.

Finally, we demonstrate experimental violation of the CHSH Bell inequality
Eq.~\eqref{eq:CHSHexp}. Experimentally, we evaluate
\begin{equation}
  C_{ij} = \frac{N^-_{ij}}{N^+_{ij}} \,,
\end{equation}
with
\begin{IEEEeqnarray}{rCl}
  N^{\pm}_{ij} &\equiv& N(\rE, \rE \mid A_{i}, B_{j}) + N(\rO, \rO \mid A_{i}, B_{j})
  \IEEEnonumber \\
  &&\pm \bigl[ N(\rE, \rO \mid A_{i}, B_{j}) + N(\rO, \rE \mid A_{i},B_{j}) \bigr] \,,
\end{IEEEeqnarray}
from the number of coincidences $N(\rE, \rE)$, $N(\rE, \rO)$, $N(\rO, \rE)$
and $N(\rO, \rO)$ simultaneously registered for each combination of outputs,
with parameters $A_{i}$ and $B_{j}$ deterministically and sequentially
selected.

Our results are shown in Table~\ref{tab:results}. One can see that the CHSH
inequality is violated by more than 40 standard deviations. Although noise is
subtracted, the theoretical optimum is not attained due to other experimental
imperfections, mainly limited visibility.

\begin{table}[!h]
  \centering
  \caption{CHSH Bell inequality violation. The first column corresponds to
    the optimal settings, $A_{i} = (a_{i}, \alpha_{i})$, $B_{j} = (b_{j},
    \beta_{j})$, $i, j = 0, 1$, computed in section \ref{sec:maxCHSH}. Second
    and third columns are theoretical predictions and experimental results,
    respectively.\label{tab:results}}
  \vspace{.25cm}
  \begin{tabular}{c r r}
    \hline & Theory & Experiment \\ \hline
    $A_0,B_0$ & $0.796$ & $0.764 \pm 0.002$  \\
    $A_0,B_1$ & $0.796$  & $0.698 \pm 0.002$  \\
    $A_1,B_0$ & $0.796$  & $0.714 \pm 0.002$  \\
    $A_1,B_1$ & $-0.178$  & $-0.158 \pm 0.002$  \\ \hline
    $S$       & $2.566$  & $2.334 \pm 0.008$ \\ \hline
  \end{tabular}
\end{table}

\section{Conclusion}

In summary, we have demonstrated by two-photon interference and Bell
inequality violation the manipulation of effective frequency qubits directly
in the frequency domain at telecommunication wavelengths using standard
telecommunication components. This further demonstrates the potential of
frequency entanglement: one has the choice to exploit high-dimensional
entanglement as in \cite{ref:oc2010,ref:om2012}, or to manipulate more
conventional two-dimensional entanglement, on which most quantum information
protocols are based.

The reported experiment could be further improved. The coincidence rate,
coincidence-to-accidental ratio, and interference visibility could be
enhanced by the use of superconducting detectors. Using a designated
filtering line and/or a source based on a resonator which would directly
produce a frequency comb of the form Eq.~\eqref{eq:psi_sep_pf} would limit
losses and enhance purity of the quantum state. The full bandwidth of the
two-photon state could be exploited provided dispersion management is
realized. These improvements would bring the method demonstrated here closer
to practical applications.

\begin{acknowledgments}
  This research was supported by the Interuniversity Attraction Poles program
  of the Belgian Science Policy Office, under grant IAP P7-35
  photonics@be, and by the F.R.S.-FNRS. E.~W. acknowledges support from
  the Belgian Fonds pour la Formation \`a la Recherche dans l'Industrie et dans
  l'Agriculture (FRIA). The authors also thank the Conseil r\'egional de
  Franche-Comt\'e and the Partenariat Hubert Curien Tournesol program.
\end{acknowledgments}

\bibliography{freq_interleavers}

\begin{thebibliography}{22}%
\makeatletter
\providecommand \@ifxundefined [1]{%
 \@ifx{#1\undefined}
}%
\providecommand \@ifnum [1]{%
 \ifnum #1\expandafter \@firstoftwo
 \else \expandafter \@secondoftwo
 \fi
}%
\providecommand \@ifx [1]{%
 \ifx #1\expandafter \@firstoftwo
 \else \expandafter \@secondoftwo
 \fi
}%
\providecommand \natexlab [1]{#1}%
\providecommand \enquote  [1]{``#1''}%
\providecommand \bibnamefont  [1]{#1}%
\providecommand \bibfnamefont [1]{#1}%
\providecommand \citenamefont [1]{#1}%
\providecommand \href@noop [0]{\@secondoftwo}%
\providecommand \href [0]{\begingroup \@sanitize@url \@href}%
\providecommand \@href[1]{\@@startlink{#1}\@@href}%
\providecommand \@@href[1]{\endgroup#1\@@endlink}%
\providecommand \@sanitize@url [0]{\catcode `\\12\catcode `\$12\catcode
  `\&12\catcode `\#12\catcode `\^12\catcode `\_12\catcode `\%12\relax}%
\providecommand \@@startlink[1]{}%
\providecommand \@@endlink[0]{}%
\providecommand \url  [0]{\begingroup\@sanitize@url \@url }%
\providecommand \@url [1]{\endgroup\@href {#1}{\urlprefix }}%
\providecommand \urlprefix  [0]{URL }%
\providecommand \Eprint [0]{\href }%
\@ifxundefined \urlstyle {%
  \providecommand \doi  [0]{\begingroup \@sanitize@url \@doi}%
  \providecommand \@doi [1]{\endgroup \@@startlink {\doibase
  #1}doi:\discretionary {}{}{}#1\@@endlink }%
}{%
  \providecommand \doi  [0]{doi:\discretionary{}{}{}\begingroup
  \urlstyle{rm}\Url }%
}%
\providecommand \doibase [0]{http://dx.doi.org/}%
\providecommand \Doi [0]{\begingroup \@sanitize@url \@Doi }%
\providecommand \@Doi  [1]{\endgroup\@@startlink{\doibase#1}\@@Doi}%
\providecommand \@@Doi [1]{#1\@@endlink}%
\providecommand \selectlanguage [0]{\@gobble}%
\providecommand \bibinfo  [0]{\@secondoftwo}%
\providecommand \bibfield  [0]{\@secondoftwo}%
\providecommand \translation [1]{[#1]}%
\providecommand \BibitemOpen [0]{}%
\providecommand \bibitemStop [0]{}%
\providecommand \bibitemNoStop [0]{.\EOS\space}%
\providecommand \EOS [0]{\spacefactor3000\relax}%
\providecommand \BibitemShut  [1]{\csname bibitem#1\endcsname}%
\bibitem [{\citenamefont {Aspect}\ \emph {et~al.}(1981)\citenamefont {Aspect},
  \citenamefont {Grangier},\ and\ \citenamefont {Roger}}]{ref:agr1981}%
  \BibitemOpen
  \bibfield  {author} {\bibinfo {author} {\bibfnamefont {A.}~\bibnamefont
  {Aspect}}, \bibinfo {author} {\bibfnamefont {P.}~\bibnamefont {Grangier}}, \
  and\ \bibinfo {author} {\bibfnamefont {G.}~\bibnamefont {Roger}},\ }\Doi
  {10.1103/PhysRevLett.47.460} {\bibfield  {journal} {\bibinfo  {journal}
  {Phys. Rev. Lett.},\ }\textbf {\bibinfo {volume} {47}},\ \bibinfo {pages}
  {460} (\bibinfo {year} {1981})}\BibitemShut {NoStop}%
\bibitem [{\citenamefont {Aspect}\ \emph {et~al.}(1982)\citenamefont {Aspect},
  \citenamefont {Grangier},\ and\ \citenamefont {Roger}}]{ref:agr1982}%
  \BibitemOpen
  \bibfield  {author} {\bibinfo {author} {\bibfnamefont {A.}~\bibnamefont
  {Aspect}}, \bibinfo {author} {\bibfnamefont {P.}~\bibnamefont {Grangier}}, \
  and\ \bibinfo {author} {\bibfnamefont {G.}~\bibnamefont {Roger}},\ }\Doi
  {10.1103/PhysRevLett.49.91} {\bibfield  {journal} {\bibinfo  {journal} {Phys.
  Rev. Lett.},\ }\textbf {\bibinfo {volume} {49}},\ \bibinfo {pages} {91}
  (\bibinfo {year} {1982})}\BibitemShut {NoStop}%
\bibitem [{\citenamefont {Kwiat}\ \emph {et~al.}(1995)\citenamefont {Kwiat},
  \citenamefont {Mattle}, \citenamefont {Weinfurter}, \citenamefont
  {Zeilinger}, \citenamefont {Sergienko},\ and\ \citenamefont
  {Shih}}]{ref:km1995}%
  \BibitemOpen
  \bibfield  {author} {\bibinfo {author} {\bibfnamefont {P.~G.}\ \bibnamefont
  {Kwiat}}, \bibinfo {author} {\bibfnamefont {K.}~\bibnamefont {Mattle}},
  \bibinfo {author} {\bibfnamefont {H.}~\bibnamefont {Weinfurter}}, \bibinfo
  {author} {\bibfnamefont {A.}~\bibnamefont {Zeilinger}}, \bibinfo {author}
  {\bibfnamefont {A.~V.}\ \bibnamefont {Sergienko}}, \ and\ \bibinfo {author}
  {\bibfnamefont {Y.}~\bibnamefont {Shih}},\ }\Doi
  {10.1103/PhysRevLett.75.4337} {\bibfield  {journal} {\bibinfo  {journal}
  {Phys. Rev. Lett.},\ }\textbf {\bibinfo {volume} {75}},\ \bibinfo {pages}
  {4337} (\bibinfo {year} {1995})}\BibitemShut {NoStop}%
\bibitem [{\citenamefont {Rarity}\ and\ \citenamefont
  {Tapster}(1990)}]{ref:rt1990}%
  \BibitemOpen
  \bibfield  {author} {\bibinfo {author} {\bibfnamefont {J.~G.}\ \bibnamefont
  {Rarity}}\ and\ \bibinfo {author} {\bibfnamefont {P.~R.}\ \bibnamefont
  {Tapster}},\ }\Doi {10.1103/PhysRevLett.64.2495} {\bibfield  {journal}
  {\bibinfo  {journal} {Phys. Rev. Lett.},\ }\textbf {\bibinfo {volume} {64}},\
  \bibinfo {pages} {2495} (\bibinfo {year} {1990})}\BibitemShut {NoStop}%
\bibitem [{\citenamefont {Mair}\ \emph {et~al.}(2001)\citenamefont {Mair},
  \citenamefont {Vaziri}, \citenamefont {Weihs},\ and\ \citenamefont
  {Zeilinger}}]{ref:mv2001}%
  \BibitemOpen
  \bibfield  {author} {\bibinfo {author} {\bibfnamefont {A.}~\bibnamefont
  {Mair}}, \bibinfo {author} {\bibfnamefont {A.}~\bibnamefont {Vaziri}},
  \bibinfo {author} {\bibfnamefont {G.}~\bibnamefont {Weihs}}, \ and\ \bibinfo
  {author} {\bibfnamefont {A.}~\bibnamefont {Zeilinger}},\ }\Doi
  {10.1038/35085529} {\bibfield  {journal} {\bibinfo  {journal} {Nature},\
  }\textbf {\bibinfo {volume} {412}},\ \bibinfo {pages} {313} (\bibinfo {year}
  {2001})}\BibitemShut {NoStop}%
\bibitem [{\citenamefont {Brendel}\ \emph {et~al.}(1992)\citenamefont
  {Brendel}, \citenamefont {Mohler},\ and\ \citenamefont
  {Martienssen}}]{ref:bmm1992}%
  \BibitemOpen
  \bibfield  {author} {\bibinfo {author} {\bibfnamefont {J.}~\bibnamefont
  {Brendel}}, \bibinfo {author} {\bibfnamefont {E.}~\bibnamefont {Mohler}}, \
  and\ \bibinfo {author} {\bibfnamefont {W.}~\bibnamefont {Martienssen}},\
  }\Doi {10.1209/0295-5075/20/7/001} {\bibfield  {journal} {\bibinfo  {journal}
  {EPL (Europhys. Lett.)},\ }\textbf {\bibinfo {volume} {20}},\ \bibinfo
  {pages} {575} (\bibinfo {year} {1992})}\BibitemShut {NoStop}%
\bibitem [{\citenamefont {Kwiat}\ \emph {et~al.}(1993)\citenamefont {Kwiat},
  \citenamefont {Steinberg},\ and\ \citenamefont {Chiao}}]{ref:ksc1993}%
  \BibitemOpen
  \bibfield  {author} {\bibinfo {author} {\bibfnamefont {P.~G.}\ \bibnamefont
  {Kwiat}}, \bibinfo {author} {\bibfnamefont {A.~M.}\ \bibnamefont
  {Steinberg}}, \ and\ \bibinfo {author} {\bibfnamefont {R.~Y.}\ \bibnamefont
  {Chiao}},\ }\Doi {10.1103/PhysRevA.47.R2472} {\bibfield  {journal} {\bibinfo
  {journal} {Phys. Rev. A},\ }\textbf {\bibinfo {volume} {47}},\ \bibinfo
  {pages} {R2472} (\bibinfo {year} {1993})}\BibitemShut {NoStop}%
\bibitem [{\citenamefont {Townsend}\ \emph {et~al.}(1993)\citenamefont
  {Townsend}, \citenamefont {Rarity},\ and\ \citenamefont
  {Tapster}}]{ref:trt1993}%
  \BibitemOpen
  \bibfield  {author} {\bibinfo {author} {\bibfnamefont {P.~D.}\ \bibnamefont
  {Townsend}}, \bibinfo {author} {\bibfnamefont {J.~G.}\ \bibnamefont
  {Rarity}}, \ and\ \bibinfo {author} {\bibfnamefont {P.~R.}\ \bibnamefont
  {Tapster}},\ }\Doi {10.1049/el:19930424} {\bibfield  {journal} {\bibinfo
  {journal} {Electron. Lett.},\ }\textbf {\bibinfo {volume} {29}},\ \bibinfo
  {pages} {634} (\bibinfo {year} {1993})}\BibitemShut {NoStop}%
\bibitem [{\citenamefont {Tittel}\ \emph {et~al.}(1998)\citenamefont {Tittel},
  \citenamefont {Brendel}, \citenamefont {Zbinden},\ and\ \citenamefont
  {Gisin}}]{ref:tb1998}%
  \BibitemOpen
  \bibfield  {author} {\bibinfo {author} {\bibfnamefont {W.}~\bibnamefont
  {Tittel}}, \bibinfo {author} {\bibfnamefont {J.}~\bibnamefont {Brendel}},
  \bibinfo {author} {\bibfnamefont {H.}~\bibnamefont {Zbinden}}, \ and\
  \bibinfo {author} {\bibfnamefont {N.}~\bibnamefont {Gisin}},\ }\Doi
  {10.1103/PhysRevLett.81.3563} {\bibfield  {journal} {\bibinfo  {journal}
  {Phys. Rev. Lett.},\ }\textbf {\bibinfo {volume} {81}},\ \bibinfo {pages}
  {3563} (\bibinfo {year} {1998})}\BibitemShut {NoStop}%
\bibitem [{\citenamefont {Tittel}\ \emph {et~al.}(2000)\citenamefont {Tittel},
  \citenamefont {Brendel}, \citenamefont {Zbinden},\ and\ \citenamefont
  {Gisin}}]{ref:tb2000}%
  \BibitemOpen
  \bibfield  {author} {\bibinfo {author} {\bibfnamefont {W.}~\bibnamefont
  {Tittel}}, \bibinfo {author} {\bibfnamefont {J.}~\bibnamefont {Brendel}},
  \bibinfo {author} {\bibfnamefont {H.}~\bibnamefont {Zbinden}}, \ and\
  \bibinfo {author} {\bibfnamefont {N.}~\bibnamefont {Gisin}},\ }\Doi
  {10.1103/PhysRevLett.84.4737} {\bibfield  {journal} {\bibinfo  {journal}
  {Phys. Rev. Lett.},\ }\textbf {\bibinfo {volume} {84}},\ \bibinfo {pages}
  {4737} (\bibinfo {year} {2000})}\BibitemShut {NoStop}%
\bibitem [{\citenamefont {Marcikic}\ \emph {et~al.}(2004)\citenamefont
  {Marcikic}, \citenamefont {de~Riedmatten}, \citenamefont {Tittel},
  \citenamefont {Zbinden}, \citenamefont {Legr\'e},\ and\ \citenamefont
  {Gisin}}]{ref:mr2004}%
  \BibitemOpen
  \bibfield  {author} {\bibinfo {author} {\bibfnamefont {I.}~\bibnamefont
  {Marcikic}}, \bibinfo {author} {\bibfnamefont {H.}~\bibnamefont
  {de~Riedmatten}}, \bibinfo {author} {\bibfnamefont {W.}~\bibnamefont
  {Tittel}}, \bibinfo {author} {\bibfnamefont {H.}~\bibnamefont {Zbinden}},
  \bibinfo {author} {\bibfnamefont {M.}~\bibnamefont {Legr\'e}}, \ and\
  \bibinfo {author} {\bibfnamefont {N.}~\bibnamefont {Gisin}},\ }\Doi
  {10.1103/PhysRevLett.93.180502} {\bibfield  {journal} {\bibinfo  {journal}
  {Phys. Rev. Lett.},\ }\textbf {\bibinfo {volume} {93}},\ \bibinfo {pages}
  {180502} (\bibinfo {year} {2004})}\BibitemShut {NoStop}%
\bibitem [{\citenamefont {Franson}(1989)}]{ref:f1989}%
  \BibitemOpen
  \bibfield  {author} {\bibinfo {author} {\bibfnamefont {J.~D.}\ \bibnamefont
  {Franson}},\ }\Doi {10.1103/PhysRevLett.62.2205} {\bibfield  {journal}
  {\bibinfo  {journal} {Phys. Rev. Lett.},\ }\textbf {\bibinfo {volume} {62}},\
  \bibinfo {pages} {2205} (\bibinfo {year} {1989})}\BibitemShut {NoStop}%
\bibitem [{\citenamefont {Ekert}\ \emph {et~al.}(1992)\citenamefont {Ekert},
  \citenamefont {Rarity}, \citenamefont {Tapster},\ and\ \citenamefont
  {Massimo~Palma}}]{ref:er1992}%
  \BibitemOpen
  \bibfield  {author} {\bibinfo {author} {\bibfnamefont {A.~K.}\ \bibnamefont
  {Ekert}}, \bibinfo {author} {\bibfnamefont {J.~G.}\ \bibnamefont {Rarity}},
  \bibinfo {author} {\bibfnamefont {P.~R.}\ \bibnamefont {Tapster}}, \ and\
  \bibinfo {author} {\bibfnamefont {G.}~\bibnamefont {Massimo~Palma}},\ }\Doi
  {10.1103/PhysRevLett.69.1293} {\bibfield  {journal} {\bibinfo  {journal}
  {Phys. Rev. Lett.},\ }\textbf {\bibinfo {volume} {69}},\ \bibinfo {pages}
  {1293} (\bibinfo {year} {1992})}\BibitemShut {NoStop}%
\bibitem [{\citenamefont {Olislager}\ \emph {et~al.}(2010)\citenamefont
  {Olislager}, \citenamefont {Cussey}, \citenamefont {Nguyen}, \citenamefont
  {Emplit}, \citenamefont {Massar}, \citenamefont {Merolla},\ and\
  \citenamefont {Phan~Huy}}]{ref:oc2010}%
  \BibitemOpen
  \bibfield  {author} {\bibinfo {author} {\bibfnamefont {L.}~\bibnamefont
  {Olislager}}, \bibinfo {author} {\bibfnamefont {J.}~\bibnamefont {Cussey}},
  \bibinfo {author} {\bibfnamefont {A.~T.}\ \bibnamefont {Nguyen}}, \bibinfo
  {author} {\bibfnamefont {P.}~\bibnamefont {Emplit}}, \bibinfo {author}
  {\bibfnamefont {S.}~\bibnamefont {Massar}}, \bibinfo {author} {\bibfnamefont
  {J.-M.}\ \bibnamefont {Merolla}}, \ and\ \bibinfo {author} {\bibfnamefont
  {K.}~\bibnamefont {Phan~Huy}},\ }\Doi {10.1103/PhysRevA.82.013804} {\bibfield
   {journal} {\bibinfo  {journal} {Phys. Rev. A},\ }\textbf {\bibinfo {volume}
  {82}},\ \bibinfo {pages} {013804} (\bibinfo {year} {2010})}\BibitemShut
  {NoStop}%
\bibitem [{\citenamefont {Olislager}\ \emph {et~al.}(2012)\citenamefont
  {Olislager}, \citenamefont {Mbodji}, \citenamefont {Woodhead}, \citenamefont
  {Cussey}, \citenamefont {Furfaro}, \citenamefont {Emplit}, \citenamefont
  {Massar}, \citenamefont {Phan~Huy},\ and\ \citenamefont
  {Merolla}}]{ref:om2012}%
  \BibitemOpen
  \bibfield  {author} {\bibinfo {author} {\bibfnamefont {L.}~\bibnamefont
  {Olislager}}, \bibinfo {author} {\bibfnamefont {I.}~\bibnamefont {Mbodji}},
  \bibinfo {author} {\bibfnamefont {E.}~\bibnamefont {Woodhead}}, \bibinfo
  {author} {\bibfnamefont {J.}~\bibnamefont {Cussey}}, \bibinfo {author}
  {\bibfnamefont {L.}~\bibnamefont {Furfaro}}, \bibinfo {author} {\bibfnamefont
  {P.}~\bibnamefont {Emplit}}, \bibinfo {author} {\bibfnamefont
  {S.}~\bibnamefont {Massar}}, \bibinfo {author} {\bibfnamefont
  {K.}~\bibnamefont {Phan~Huy}}, \ and\ \bibinfo {author} {\bibfnamefont
  {J.-M.}\ \bibnamefont {Merolla}},\ }\Doi {10.1088/1367-2630/14/4/043015}
  {\bibfield  {journal} {\bibinfo  {journal} {New J. Phys.},\ }\textbf
  {\bibinfo {volume} {14}},\ \bibinfo {pages} {043015} (\bibinfo {year}
  {2012})}\BibitemShut {NoStop}%
\bibitem [{\citenamefont {Merolla}\ \emph
  {et~al.}(1999){\natexlab{a}}\citenamefont {Merolla}, \citenamefont
  {Mazurenko}, \citenamefont {Goedgebuer}, \citenamefont {Porte},\ and\
  \citenamefont {Rhodes}}]{ref:mm1999a}%
  \BibitemOpen
  \bibfield  {author} {\bibinfo {author} {\bibfnamefont {J.-M.}\ \bibnamefont
  {Merolla}}, \bibinfo {author} {\bibfnamefont {Y.}~\bibnamefont {Mazurenko}},
  \bibinfo {author} {\bibfnamefont {J.-P.}\ \bibnamefont {Goedgebuer}},
  \bibinfo {author} {\bibfnamefont {H.}~\bibnamefont {Porte}}, \ and\ \bibinfo
  {author} {\bibfnamefont {W.~T.}\ \bibnamefont {Rhodes}},\ }\Doi
  {10.1364/OL.24.000104} {\bibfield  {journal} {\bibinfo  {journal} {Opt.
  Lett.},\ }\textbf {\bibinfo {volume} {24}},\ \bibinfo {pages} {104} (\bibinfo
  {year} {1999}{\natexlab{a}})}\BibitemShut {NoStop}%
\bibitem [{\citenamefont {Merolla}\ \emph
  {et~al.}(1999){\natexlab{b}}\citenamefont {Merolla}, \citenamefont
  {Mazurenko}, \citenamefont {Goedgebuer},\ and\ \citenamefont
  {Rhodes}}]{ref:mm1999b}%
  \BibitemOpen
  \bibfield  {author} {\bibinfo {author} {\bibfnamefont {J.-M.}\ \bibnamefont
  {Merolla}}, \bibinfo {author} {\bibfnamefont {Y.}~\bibnamefont {Mazurenko}},
  \bibinfo {author} {\bibfnamefont {J.-P.}\ \bibnamefont {Goedgebuer}}, \ and\
  \bibinfo {author} {\bibfnamefont {W.~T.}\ \bibnamefont {Rhodes}},\ }\Doi
  {10.1103/PhysRevLett.82.1656} {\bibfield  {journal} {\bibinfo  {journal}
  {Phys. Rev. Lett.},\ }\textbf {\bibinfo {volume} {82}},\ \bibinfo {pages}
  {1656} (\bibinfo {year} {1999}{\natexlab{b}})}\BibitemShut {NoStop}%
\bibitem [{\citenamefont {Bloch}\ \emph {et~al.}(2007)\citenamefont {Bloch},
  \citenamefont {McLaughlin}, \citenamefont {Merolla},\ and\ \citenamefont
  {Patois}}]{ref:bm2007}%
  \BibitemOpen
  \bibfield  {author} {\bibinfo {author} {\bibfnamefont {M.}~\bibnamefont
  {Bloch}}, \bibinfo {author} {\bibfnamefont {S.~W.}\ \bibnamefont
  {McLaughlin}}, \bibinfo {author} {\bibfnamefont {J.-M.}\ \bibnamefont
  {Merolla}}, \ and\ \bibinfo {author} {\bibfnamefont {F.}~\bibnamefont
  {Patois}},\ }\Doi {10.1364/OL.32.000301} {\bibfield  {journal} {\bibinfo
  {journal} {Opt. Lett.},\ }\textbf {\bibinfo {volume} {32}},\ \bibinfo {pages}
  {301} (\bibinfo {year} {2007})}\BibitemShut {NoStop}%
\bibitem [{\citenamefont {Bennett}\ and\ \citenamefont
  {Brassard}(1984)}]{ref:bb1984}%
  \BibitemOpen
  \bibfield  {author} {\bibinfo {author} {\bibfnamefont {C.~H.}\ \bibnamefont
  {Bennett}}\ and\ \bibinfo {author} {\bibfnamefont {G.}~\bibnamefont
  {Brassard}},\ }in\ \href@noop {} {\emph {\bibinfo {booktitle} {Proceedings of
  IEEE International Conference on Computers, Systems and Signal Processing}}}\
  (\bibinfo  {publisher} {IEEE},\ \bibinfo {address} {New York},\ \bibinfo
  {year} {1984})\ pp.\ \bibinfo {pages} {175--179}\BibitemShut {NoStop}%
\bibitem [{\citenamefont {Clauser}\ \emph {et~al.}(1969)\citenamefont
  {Clauser}, \citenamefont {Horne}, \citenamefont {Shimony},\ and\
  \citenamefont {Holt}}]{ref:ch1969}%
  \BibitemOpen
  \bibfield  {author} {\bibinfo {author} {\bibfnamefont {J.~F.}\ \bibnamefont
  {Clauser}}, \bibinfo {author} {\bibfnamefont {M.~A.}\ \bibnamefont {Horne}},
  \bibinfo {author} {\bibfnamefont {A.}~\bibnamefont {Shimony}}, \ and\
  \bibinfo {author} {\bibfnamefont {R.~A.}\ \bibnamefont {Holt}},\ }\Doi
  {10.1103/PhysRevLett.23.880} {\bibfield  {journal} {\bibinfo  {journal}
  {Phys. Rev. Lett.},\ }\textbf {\bibinfo {volume} {23}},\ \bibinfo {pages}
  {880} (\bibinfo {year} {1969})}\BibitemShut {NoStop}%
\bibitem [{\citenamefont {Clauser}\ and\ \citenamefont
  {Horne}(1974)}]{ref:ch1974}%
  \BibitemOpen
  \bibfield  {author} {\bibinfo {author} {\bibfnamefont {J.~F.}\ \bibnamefont
  {Clauser}}\ and\ \bibinfo {author} {\bibfnamefont {M.~A.}\ \bibnamefont
  {Horne}},\ }\Doi {10.1103/PhysRevD.10.526} {\bibfield  {journal} {\bibinfo
  {journal} {Phys. Rev. D},\ }\textbf {\bibinfo {volume} {10}},\ \bibinfo
  {pages} {526} (\bibinfo {year} {1974})}\BibitemShut {NoStop}%
\bibitem [{\citenamefont {Cuyt}\ \emph {et~al.}(2008)\citenamefont {Cuyt},
  \citenamefont {Brevik~Petersen}, \citenamefont {Verdonk}, \citenamefont
  {Waadeland},\ and\ \citenamefont {Jones}}]{ref:cbp2008}%
  \BibitemOpen
  \bibfield  {author} {\bibinfo {author} {\bibfnamefont {A.}~\bibnamefont
  {Cuyt}}, \bibinfo {author} {\bibfnamefont {V.}~\bibnamefont
  {Brevik~Petersen}}, \bibinfo {author} {\bibfnamefont {B.}~\bibnamefont
  {Verdonk}}, \bibinfo {author} {\bibfnamefont {H.}~\bibnamefont {Waadeland}},
  \ and\ \bibinfo {author} {\bibfnamefont {W.~B.}\ \bibnamefont {Jones}},\
  }\enquote {\bibinfo {title} {Handbook of continued fractions for special
  functions},}\ \ (\bibinfo  {publisher} {Springer Netherlands},\ \bibinfo
  {year} {2008})\ p.\ \bibinfo {pages} {344},\ ISBN \bibinfo {isbn}
  {9781402069482}\BibitemShut {NoStop}%
\end{thebibliography}%

\end{document}